\documentstyle[twocolumn,aps,prl,graphics,epsfig,floats]{revtex}
\begin{document}
\draft
\twocolumn
\wideabs{
\title{Entangled state quantum cryptography: 
Eavesdropping on the Ekert protocol}
\vskip -0.8 cm
\author{D. S. Naik$^{1}$, C. G. Peterson$^{1}$, A. G. White$^{1,2}$,
A. J. Berglund$^{1}$, and P. G. Kwiat$^{1*}$
}
\address{$^{1}$ Physics Division, P-23, Los Alamos National Laboratory,
Los Alamos, New Mexico 87545, USA}
\address{$^{2}$ Department of Physics, University of Queensland,
Brisbane, Queensland 4072, AUSTRALIA}
\date{Submitted to Phys. Rev. Lett, Oct. 17, 1999; resubmitted 
12/22/99}

\maketitle
\begin{abstract}
Using polarization-entangled photons from spontaneous
parametric downconversion, we have implemented Ekert's quantum
cryptography protocol.  The near-perfect correlations of the photons 
allow the sharing of a secret
key between two parties.  The presence of an eavesdropper is
continually checked by measuring Bell's inequalities.  We investigated
several possible eavesdropper strategies, including pseudo-quantum
non-demolition measurements.  In all cases, the eavesdropper's presence 
was readily apparent.  We discuss a procedure to increase her detectability.

\end{abstract}
\draft
}\narrowtext
The emerging field of quantum information science
aims to use the nonclassical features of quantum systems to achieve
performance in communications and computation that is superior to that
achievable with systems based solely on classical physics.  For 
example, current methods 
of public-key cryptography base their security on the supposed
(but unproven) computational difficulty in solving certain problems,
e.g., finding the prime factors of large numbers -- these problems have
not only been unproven to be difficult, but actually been shown to be
computational ``easy'' in the context of quantum computation
\cite{Shor}.  In contrast, it is now generally accepted that 
techniques of quantum
cryptography can allow completely secure communications between
distant parties \cite{BB84,Ekert,Bennett,B92}.  
Specifically, by using single
photons to distribute a secret random cryptographic key, one can
ensure that no eavesdropping goes unnoticed; more precisely, one can
set rigid upper bounds on the possible information known to a
potential eavesdropper, based on measured error rates, and 
use appropriate methods of ``privacy amplification'' to reduce this
information to an acceptable level \cite{privacy}.

Since its discovery, quantum cryptography has been demonstrated by a
number of groups using weak coherent states, both in fiber-based
systems \cite{fiber_exps} and in free space arrangements
\cite{Bennett_exp,freespace}.  These experiments are 
provably secure against all eavesdropping attacks based 
on presently available technology; however, there are 
certain conceivable attacks to which they are 
vulnerable, as sometimes the pulses used necessarily 
contain more than one photon -- an eavesdropper could in 
principle use these events to gain information 
about the key without introducing any extra errors \cite{loophole}.   
Use of true single-photon sources can close this potential 
security loophole; and while the loophole still exists when 
using {\it pairs} of photons as from parametric down-conversion 
(because occasionally there will be 
{\it double} pairs), it has been shown that they may allow secure 
transmissions over longer distances \cite{Lutkenhaus}.

While a number of groups use correlated photon pairs to study nonlocal
correlations (via tests of Bell's inequalities 
\cite{Bell,Bellexps,Kwiat}), and their
possible application for quantum cryptography \cite{Gisin,Sasha}, to
our knowledge no results explicitly using entangled photons in a
quantum cryptographic protocol have been reported in the
literature\cite{Jennewein}.  It is now well established that one
cannot employ these nonlocal correlations 
for superluminal signaling\cite{eberhard}.  
Nevertheless, Ekert showed that one can use the
correlations to establish a secret random key between two parties, as
part of a completely secure cryptography protocol \cite{Ekert}.

In our version of the Ekert protocol, ``Alice'' and ``Bob'' each
receive one photon of a polarization-entangled pair in the state
$|\phi^{+}\rangle = (|H_{1}H_{2}\rangle +
|V_{1}V_{2}\rangle)/\sqrt{2}$, where $H$ ($V$) represents horizontal
(vertical) polarization.  They each respectively measure the
polarization of their photons in the bases $(|H_{1}\rangle + e^{i
\alpha}|V_{1}\rangle)$ and $(|H_{2}\rangle + e^{i
\beta}|V_{2}\rangle)$, where $\alpha$ and $\beta$ randomly take on the
values $\alpha_1\!=\!45^{\circ}, \alpha_2\!=\!90^{\circ},
\alpha_3\!=\!135^{\circ}, \alpha_4\!=\!180^{\circ};
\beta_1\!=\!0^{\circ}, \beta_2\!=\!45^{\circ}, \beta_3\!=\!90^{\circ},
\beta_4\!=\!135^{\circ}$.  
They then disclose by public discussion which bases were used, but
not the measurement results.
For the state $|\phi^{+}\rangle$, the
probabilities for a coincidence between Alice's detector 1 (or
detector 1', which detects the orthogonally-polarized photons) and
Bob's detector 2 (2') are given by
\begin{eqnarray}
\rm{P}_{12}(\alpha,\beta)
     &=&
\rm{P}_{1'2'}(\alpha,\beta)
     = (1 + cos(\alpha + \beta))/4\\
\rm{P}_{12'}(\alpha,\beta)
     &=&
\rm{P}_{1'2}(\alpha,\beta)
     = (1 - cos(\alpha + \beta))/4 \,\,.\nonumber
\end{eqnarray}
Note that 
when $\alpha + \beta= 180^{\circ}$, 
they will have completely correlated results, which
then constitute the quantum cryptographic key.  As indicated in 
Table~\ref{tab:settings}, the results from other 
combinations are revealed and 
used in two independent tests of Bell's inequalities, to
check the presence of an intermediate eavesdropper (``Eve'').  Here we 
\begin{table}[!b]
\caption{Distribution of data dependent on Alice's and Bob's
respective phase settings $\alpha_{i}$ and $\beta_{i}$ (see text
for details).} 
\begin{center}
\begin{tabular}{cc| c c c c }
             &   		&              &      Alice        & &   \\*
  &           		& $\alpha_{1}$ & $\alpha_{2}$ & $\alpha_{3}$
&  $\alpha_{4}$  \\* \hline
            &$\beta_{1}$	&S		& --		& S
		& QKey \\*
     Bob    &$\beta_{2}$	&--		& $S'$		& QKey
		& $S'$ \\*
            &$\beta_{3}$	&S		&QKey		& S
		& -- \\*
            &$\beta_{4}$	&QKey		& $S'$		& --
		& $S'$ \\*
\end{tabular}
\end{center}
\label{tab:settings}
\end{table}
\noindent
\begin{figure}[t!]
    \vspace {-0.3 cm}
\begin{center}
\epsfxsize=\columnwidth
\epsfbox{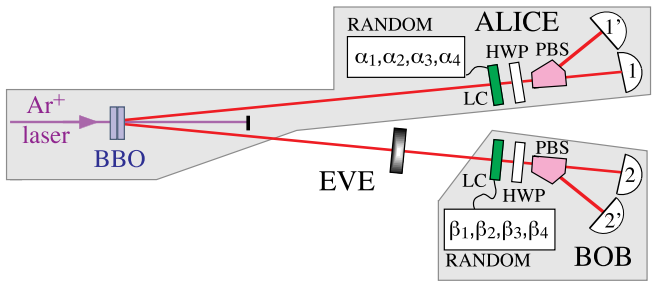}
\end{center}
\caption{Schematic of quantum cryptography system.  351.1nm light from
an Argon ion laser is used to pump two perpendicularly-oriented
nonlinear optical crystals (BBO).  The resultant entangled photons are
sent to Alice and Bob, who each analyze them in one of four randomly
chosen bases.  The eavesdropper, if present, was incorporated using
either a polarizer or a decohering birefringent plate (both
orientable, and in some cases with additional wave plates to allow
analysis in arbitrary elliptical polarization bases [Fig.  2a,c]).}
\label{fig:setup}
\end{figure}
\noindent 
present an experimental realization of this protocol, and look at the
effect of various eavesdropping strategies.

We prepare the polarization-entangled state using the process of
spontaneous parametric down-conversion 
in a nonlinear crystal\cite{Kwiat}.  In brief, 
two identically-cut adjacent crystals
(beta-barium-borate, BBO) are oriented with their optic axes in planes
perpendicular to each other (Fig.  1).  A $45^{\circ}$-polarized pump
photon is then equally likely to convert in either crystal.  Given the
coherence and high spatial overlap (for our 0.6mm-thick crystals)
between these two processes, the photon pairs are then created in the
maximally-entangled state $|\phi^{+}\rangle$.  Alice's and Bob's
analysis systems each consist of a randomly driven liquid crystal [LC]
(to set the applied phase shift), a half wave plate [HWP] (with optic
axis at $22.5^{\circ}$), and a calcite Glan-Thompson prism [PBS].
Photons from the horizontal and vertical polarization outputs of each
prism are detected (after narrow-band interference filters) using
silicon avalanche photodiodes (EG\&G SPCM-AQ's, efficiency $\sim
60\%$, dark count $< 400s^{-1}$).  The correlated detector signals are
synchronized and temporally discriminated through AND gates.  Because
of the narrow gate window (5 ns), the rate of accidental coincidences
(resulting from multiple pairs or background counts) is only
$10^{-5}s^{-1}$.  From separate computers, Alice and Bob control their
respective LC's with synchronously clocked arbitrary waveform
generator cards \cite{random}.  A coincident event triggers a
digitizer, which records the LC states, and the outputs from each of
the four detector pairs \cite{key}.

Because the total rate of coincidences between Alice's and Bob's
detectors was typically 5000 s$^{-1}$, the probability of having at
least one pair of photons during the collection time window of 1 ms was
99\%.  Of course, there was then also a high probability of more than
one pair being detected within the window (96\%).  Because the phase
setting remains unchanged during a collection window, muliple pairs
could conceivably give extra information to a potential eavesdropper.
We avoided this problem by keeping 
only the first event in any given window.
Assuming that Alice and Bob each have completely isolated measurement
systems (i.e., there is no way for an eavesdropper to learn about the
measurement parameters $\alpha$ and $\beta$ by sending in extra
photons of her own), this system is secure even though no rapid
switching is employed, since only $\sim$1 photon pair event is used for
any particular $\alpha$-$\beta$ setting \cite{doubles}.  
Given the 22ms cycle period
determined by the liquid crystals \cite{liquidcrystals}, the maximum
rate of data collection in our system is 45.4Hz.  The usable rate is
slightly less, because the LC voltages were occasionally in transition
when the digitizer read them, yielding an ambiguous determination of
the actual phase setting.  Typically we collected 40 useful pairs per
second.

As seen in Table I, only 1/4 of the data actually contributes to the
raw cryptographic key; half the data are used to test Bell's inequalities;
and 1/4 are not used at all \cite{threephases}.  In four independent
runs of $\sim$10 minutes each, we obtained a total of 24252 secret key
bits (see Table II), corresponding to a raw bit rate of 10.1s$^{-1}$; 
the corresponding bit error rate (BER) was $3.06\pm 0.11\%$\cite{BER}.  If 
we attribute this BER (conservatively estimated as 3.4\%) 
entirely to an eavesdropper, we
should assume she has knowledge of up to 0.7\% + (4/$\sqrt(2)* 
3.4\% = 10.3\%$ ($\sim$2500 bits) of the key, 
where the 0.7\% comes from possible double-pair 
events \cite{doubles}, and the second term assumes an intercept-resend
strategy (see \cite{Bennett_exp}).  We must then perform sufficient privacy
amplification to reduce this to an acceptable level.
After running an error detection procedure on our raw key material,
17452 error-free bits remained.  Using appropriate privacy
amplification techniques \cite{privacy}, this was further compressed
to 12215 useful secret bits (a net bit rate of 5.1s$^{-1}$); 
the residual information available to any
potential eavesdropper is then $2^{-(17452-12215-2500)}/\ln 2$, i.e.,
$\ll 1$ bit \cite{Bennett_exp}.

In contrast to nearly all tests of Bell's inequalities previously
reported, instead of using linear polarization
analyses (i.e., in the equatorial plane of the Poincar\'{e} sphere),
we used elliptical polarization analysis (i.e., on the plane
containing the circularly polarized poles of the sphere and the $\pm
45^{\circ}$ linearly polarized states).  In particular, we measured
the Bell parameters\cite{CHSH}:
\begin{eqnarray}
\rm{S} = -E(\alpha_{1},\beta_{1})
+E(\alpha_{1},\beta_{3})+E(\alpha_{3},
\beta_{1})+E(\alpha_{3},\beta_{3})\\
S'\rm{} = E(\alpha_{2},\beta_{2})
        +E(\alpha_{2},\beta_{4})+E(\alpha_{4},\beta_{2})
-E(\alpha_{4},\beta_{4})\;,\nonumber
\end{eqnarray}
where
E($\alpha,\!\beta$) = $\frac{\rm{R}_{12}
(\alpha,\!\beta)+\rm{R}_{1'2'}(\alpha,\!\beta)
        -\rm{R}_{12'}(\alpha,\!\beta)-\rm{R}_{1'2}(\alpha,\!\beta)}
        {\rm{R}_{12}(\alpha,\!\beta)+\rm{R}_{1'2'}(\alpha,\!\beta)
        +\rm{R}_{12'}(\alpha,\!\beta)+\rm{R}_{1'2}(\alpha,\!\beta)}$,
and the R's are the various coincidence counts between Alice's and
Bob's detectors.  For any local realistic theory $|$S$|$,$|S'|\le2$,
while for the combinations of $\alpha$ and $\beta$ 
indicated in Table 
1, the quantum mechanically expected values of $|$S$|$,$|S'|$ are
$2\sqrt{2}$.  In a typical 10 minute run of our system, we observed
S=$-2.67\pm 0.04$ and $S'$=$-2.65\pm 0.04$; for the 40 minutes of
collected data, our combined values were S=$-2.665\pm 0.019$,
$S'$=$-2.644\pm 0.019$, each a 34-$\sigma$ violation of Bell's
inequality.  It is expected (and demonstrated experimentally; see
below) that the presence of an eavesdropper will reduce the observed
values of $|$S$|$,$|S'|$.  In fact, if the eavesdropper measures one
photon from every pair, then $|S_{eve}| \le \sqrt{2}$\cite{Ekert}.   
Because we observed high values of $|$S$|$,$|S'|$,
in our system the presence of an eavesdropper could thus be detected
in $\sim$1s of data collection (the time interval for which our
$|$S$|$,$|S'|$ exceed $\sqrt{2}$ by 2$\sigma$).  Of course one could
similarly use the BER as a check for a potential eavesdropper, who
introduces a minimum BER of 25\% if she measures every photon;
however, this requires sacrificing some of the cryptographic key to
accurately determine the BER.
\begin{table}[!b]
    \vspace {-0.1cm}
\caption{100 bits of typical shared quantum key data for Alice (A) and
Bob (B), generated using the Ekert protocol.  Italic entries indicate
errors; our average BER was 3.06\%.}
\begin{center}
\begin{tabular}{c  }
A: 1111100101010110100110000 1010011100110{\it 1}11010100000\\*
B: 1111100101010110100110000 1010011100100{\it 0}11010101000\\*
\hline
A: 100010010100001{\it 0}100111101 110100100101{\it 0}101010010111\\*
B: 110010010100001{\it 1}100111101 110100100101{\it 1}101010010111\\*
\end{tabular}
\end{center}
\label{tab:data}
\vspace {-0.3cm}
\end{table}
\noindent 

In investigating the effects of the presence of an eavesdropper there
are two main difficulties.  First, there are various possible strategies; and
second, we always assume that Eve has essentially perfect equipment
and procedures, 
which of course is experimentally impossible to
implement.  Hence, we can at best simulate the effects she would
have; we did this for two particular intercept/resend
eavesdropping strategies.  In the first, we make a
strong filtering measurement of the polarization, in some basis $\chi$, and
send on the surviving photons to Bob.  The simulated 
eavesdropper thus 
makes the projective measurement $|\chi\rangle\langle\chi|$. The
effect on the measured value of S and $S'$ and the BER depend strongly
on what eavesdropping basis $|\chi\rangle$ is used 
\cite{Bennett_exp}.
Theoretical predictions and results for bases in three
orthogonal planes in the Poincar{\'e} sphere are shown in Figure 2.

The second eavesdropping strategy examined was a quantum
non-demolition (QND) measurement \cite{Werner}.  
QND measurements of {\it optical}
photon number and polarization are presently impossible.  
In fact, precisely for this reason current quantum
cryptography implementations {\it are} secure, even though they employ
weak optical pulses (with {\it average} photon number/pulse less than
1) \cite{Durt_reply}.  Nevertheless, the ideal of quantum cryptography
is that it can be made secure against {\it any} physically possible
eavesdropping strategies; hence, it is desirable to test any system
against as many strategies as possible.

Although appropriate QND measurements cannot be performed at present,
it is well known that their {\it effect} is to produce a random phase
between the eigenstates of the measurement, in turn due to the
entanglement of these states with the readout quantum system.  We can
exactly simulate this effect by inserting, in Bob's path, a
birefringent element that separates the extraordinary and ordinary
components of the photon wavepacket by more than the coherence length
($\sim 140\mu$m, determined by the interference filters before the
detectors); the result is a completely random phase between these
polarization 
\begin{figure}[t!]
    \vspace {-0.2 cm}
\begin{center}
\epsfxsize=\columnwidth
\epsfbox{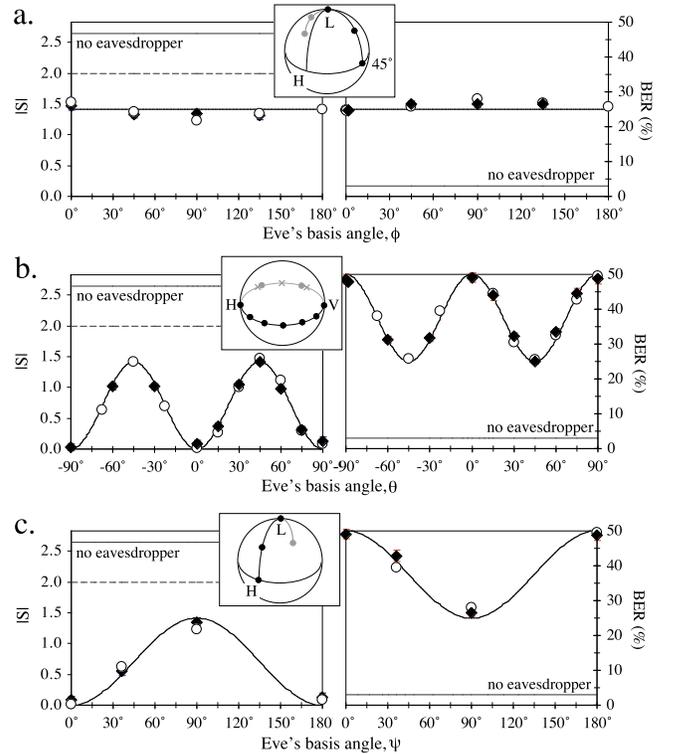}
\end{center}
\caption{Data and theory showing the effect of an eavesdropper on S
and BER for various attack bases (as $S'$ closely agrees with S, it is
omitted for clarity).  Diamonds represent strong measurements, made
with a polarizer; circles represent QND attacks, simulated with a
3mm-thick BBO crystal; error bars are within the points.  The attacks
bases are: a.  $|H\rangle + e^{i\phi}|V\rangle$; b.
$\cos\theta|H\rangle + \sin\theta|V\rangle$; and c.
$|45^{\circ}\rangle + e^{i\psi}|$-$45^{\circ}\rangle$; the actual
measurement points in these bases are illustrated on the inset
Poincar{\'e} spheres.  The measured average values with no
eavesdropper are indicated by unbroken grey lines, the broken lines
represent the maximum classical value of $|$S$|$.}
    \vspace {-0.1 cm}

\end{figure}
\noindent
components, just as if a QND measurement had been made on
them.  Mathematically, the effect of the eavesdropper is to make a
projective measurement $|\chi\rangle\langle\chi|+ e^{i
\langle\xi\rangle} |\chi^{\perp}\rangle\langle\chi^{\perp}|$, where
$\langle\xi\rangle$ represents a random phase.  Note that the
theoretical predictions are identical with that for the strong
polarization measurement.  The experimental data are also shown in
Fig. 2.

We see immediately that the optimal bases for eavesdropping lie in the
same plane (on the Poincar{\'e} sphere) as the bases employed by Alice
and Bob -- for this case, the probability that the eavesdropper causes
an error is ``only'' 25\% per intercepted bit, and the $|$S$|$ value
is $\sqrt{2}$ (Fig.  2a).  On the other hand, if the eavesdropper does
not know the plane of the measurement bases, and uses, e.g., random
measurements in an orthogonal plane, her {\it average} probability of
producing an error climbs to 32.5\%/bit, and the average value of
$|$S$|$ drops to $1/\sqrt{2}$.  This suggests a strategy for improved
security: 
Alice and Bob should choose bases corresponding to at least two (and
ideally three) orthogonal planes, thereby ``magnifying'' the presence of an
eavesdropper (at least one implementing the sort of strong projective
or QND-like measurements strategies investigated here) 
above the usual 25\%/bit error probability.  Quantitative 
theoretical investigations of such a strategy, known as the ``six-state''
protocol, support these claims \cite{Bruss}.

An eavesdropper could also examine only a {\it fraction} 
of the photons, thus reducing her induced BER and increasing
the S value measured by Alice and Bob, at the expense of her own
knowledge of the cryptographic key.  For example, if she measures (in
the optimal basis) less than 58.6\% of the photons, S $> 2$ and the
corresponding BER $< 15$\%, but Eve's knowledge of the key will be
less than Bob's (and privacy amplification techniques will
still permit generation of a secret key) 
\cite{Gisin2,11limit}.

In summary, we have implemented the Ekert quantum cryptography
protocol using entangled photon pairs.  For this proof-of-principle
experiment, Alice and Bob were situated side by side on the same
optical table, obviously not the optimal configuration for useful
cryptography.  Nevertheless, our system demonstrates the essential
features of the Ekert protocol, and moreover, we believe is the first
to {\it experimentally} investigate the effect of a physical
intermediate eavesdropper \cite{exp_eve}.  We see no bar to extending
the transmission distance to hundreds of meters \cite{freespace} or
even to earth-to-satellite distances \cite{Richard}.  

We gratefully acknowledge the laboratory assistance of S. Lopez, the
error correction/privacy amplification programs written by E.
Twyeffort, and very helpful discussions with R. Hughes and N. 
Lutkenhaus.  

$^{*}$Please address correspondence to: Kwiat@lanl.gov.


\begin{references}
\bibitem{random} The random pulse sequences to drive the LC's were
generated by LabView on each of Alice's and Bob's separate computers.
For our comparatively short data strings, the pseudo-random numbers were
adequate.
\bibitem{doubles}
There is a small contribution ($\sim$ 0.7\% per key bit) of events 
in which a double pair was emitted within the detection time 
(conservatively estimated as the 5ns gate window + 35ns dead time).  
In principle an eavesdropper could learn the value of these key bits, 
using methods similar to those for weak pulse schemes \cite{loophole}.
\bibitem{liquidcrystals} Heating the LC's to $35^{\circ}$, and
applying ``overshoot'' voltages, improved the switching times to under
19 ms for all transitions; photons were collected in the following 1
ms.  The cycle time could be improved to nanoseconds by using, e.g.,
electro-optic modulators instead of LC's.
\bibitem{BER} The BER is defined as the number of errors divided by
the total size of the cryptographic key; BER = 0.5 implies Alice and
Bob have completely uncorrelated strings.
\end{references}
\end{document}